\documentclass[12pt,preprint]{aastex63}
\usepackage{epstopdf}
\usepackage{natbib}
\usepackage{graphicx}
\usepackage{CJK}

\newcommand       \RV           {R_{\rm V}}
\newcommand       \AV           {A_{\rm V}}
\newcommand       \Teff         {T_{\rm {eff}}}
\accepted{April 16, 2021}
\submitjournal{ApJS}
\begin{document}
\begin{CJK*}{UTF8}{gbsn}
\title{The Ultraviolet Extinction Map and Dust Property at High Galactic Latitude}
\author[0000-0002-2473-9948]{Mingxu Sun (孙明旭)}
\affiliation{Department of Astronomy,
               Beijing Normal University,
               Beijing 100875, China}
\author[0000-0003-3168-2617]{Biwei Jiang(姜碧沩)}
\affiliation{Department of Astronomy,
               Beijing Normal University,
               Beijing 100875, China}
\author[0000-0003-2471-2363]{Haibo Yuan(苑海波)}
\affiliation{Department of Astronomy,
               Beijing Normal University,
               Beijing 100875, China}
\author[0000-0001-9328-4302]{Jun Li (李军)}
\affiliation{Department of Astronomy,
               Beijing Normal University,
               Beijing 100875, China}

\correspondingauthor{Biwei Jiang}
\email{bjiang@bnu.edu.cn}

\begin{abstract}

Extinction in ultraviolet is much more significant than  in optical or infrared, which can be very informative to  precisely measure the extinction and understand the dust properties in the low extinction areas. The high Galactic latitude sky is such an area, important for studying the extragalactic sky and the universe. Based on the stellar parameters measured by the LAMOST and GALAH spectroscopy and the ultraviolet photomery by the \emph{GALEX} space telescope, the extinction of 1,244,504 stars in the \emph{GALEX}/NUV band and 56,123 stars in the \emph{GALEX}/FUV band are calculated precisely. The error of color excess is 0.009, 0.128 and 0.454 mag for $E_{\rm G_{BP}, G_{RP}}$, $E_{\rm NUV,G_{BP}}$ and $E_{\rm FUV,G_{BP}}$ respectively. They delineates the \emph{GALEX}/NUV extinction map of about a third of the sky mainly at the high Galactic latitude area with an angular resolution of $\sim 0.4\,\, \rm deg$. The mean color excess ratio in the entire sky areas is derived to be 3.25, 2.95 and -0.37 for $E_{{\rm NUV,G_{BP}}} / E_{{\rm G_{BP},G_{RP}}}$, $E_{{\rm FUV,G_{BP}}} / E_{{\rm G_{BP},G_{RP}}}$ and $E_{{\rm FUV,NUV}} / E_{{\rm G_{BP},G_{RP}}}$ respectively, which is in general agreement with the previous works, and their changes with the Galactic latitude and the interstellar extinction are discussed.


\end{abstract}


\section{Introduction}

The interstellar medium is full of dust and gas, where the dust, though about one percent in mass, is the dominant factor in extinction. The interstellar extinction changes from sightline to sightline and from distance to distance, even the extinction law itself changes with environment (see e.g. \citet{2003ARA&A..41..241D}). Because interstellar dust absorbs and scatters light that reforms the appearance of objects, a precise determination of interstellar extinction map is  crucial to recover the true brightness and spectrum of the objects. Thus there have been many hard efforts to construct the extinction map.

The general two-dimensional (2D)  extinction map usually yields  the integral extinction along a given sightline, which can help make a rough correction without the distance information. \citet[SFD98]{1998ApJ...500..525S}  constructed a 2D reddening map based on the far-infrared emission of dust. Assuming a modified blackbody emission law in far-infrared as well as a linear relationship between the dust optical depth at 100\,$\rm {\mu m}$ and the reddening, they calibrate the optical depth map according to the measured color excess value of $E(B-V)$.  Later, \citet{2010ApJ...719..415P} provide a correction to the SFD98 extinction map by using passively evolving galaxies as color standards to measure deviations. The maps have an angular resolution of $4.5^{\circ}$ and a 1$\sigma$ uncertainty of 1.5 mmag in $E(B-V)$, which are largely accurate for most of the areas with the $E(B-V)$ deviation below 3\,mmag, but some regions deviate from SFD98 by as much as 50\%. Spatial structure of the deviation is strongly correlated with the observed dust temperature in that SFD98 underpredict reddening in the regions of low dust temperature. With the newly available \emph{PLANCK} all-sky observation, \citet{2014A&A...571A..11P} also constructed a  2D map of dust reddening across the entire sky. They fit a modified blackbody dust emission model to the \emph{Planck} and \emph{IRAS} far-infrared emission maps, and convert the dust emission into interstellar reddening from three parameters: dust optical depth at 353 GHz, thermal dust radiance and a recommended extragalactic reddening estimate with point sources removed. Afterwards, \citet{2016A&A...586A.132P} present an all-sky modelling of \emph{Planck}, \emph{IRAS}, and \emph{WISE} observations using the physical dust model by \citet[DL07]{2007ApJ...657..810D}. They employ the DL07 dust model to generate maps of the dust mass surface density, the dust optical extinction etc. While these 2D maps are all based on the dust emission in far-infrared, recently some 3D extinction maps are obtained with the distance measured by the \emph{Gaia} mission. With the availability of \emph{Gaia}/DR2 \citep{2018A&A...616A...1G}, \citet{2018A&A...616A...8A}, \citet{2019MNRAS.483.4277C} and \citet{2019ApJ...887...93G} established the 3D interstellar-reddening map, which are derived from the multi-band photometries of hundreds of millions of stars.

However, all these extinction maps are based on the infrared and optical observation, which may convert to the UV extinction from the given visual/infrared extinction. \citet[CCM89]{1989ApJ...345..245C}, \citet[O'D94]{1994ApJ...422..158O} and \citet[F99]{1999PASP..111...63F}  all report reddening laws with a single parameter, $\RV \equiv \AV/ E(B-V)$, which implies some conversion factor from the visual to the UV extinction.  Nevertheless, these laws are derived by the relatively large-extinction stars. The vast areas of the sky of much lower extinction have not been adequately examined, and  extrapolating these extinction laws to yield the UV extinction at high latitudes may not be that convincing. According to the variation in the UV colors and number density of galaxies, \cite{2013ApJ...771...68P} present an analysis of 373,303 galaxies selected from the \emph{GALEX} and \emph{ WISE} catalogs. They find that dust at high latitude is neither quantitatively nor qualitatively consistent with standard reddening laws. Extinction in the $FUV$ and $NUV$ is about 10\% and 35\% higher than expected, with significant variation across the sky. They find that no single $\RV$ parameter fits both the optical and ultraviolet extinction at high latitude, and that while both show detectable variation across the sky, these variations are not related.
The ultraviolet (UV) extinction map needs to be constructed separately considering the aforementioned variations. In comparison with optical and infrared wavelengths, the UV bands are more sensitive to interstellar extinction. For an average $\RV$=3.1 law, the extinction at 2000\,{\AA} is about three times that at the visual 5000\,{\AA} \citep{1990ARA&A..28...37M}. Thus the UV extinction becomes evident at the low-extinction regions where the visual/infrared extinction may be very small or undetectable. One such region is at the high Galactic latitude important for the study of the extragalactic sky and cosmology, where the UV extinction is of great significance to correct the extinction precisely. In addition, the UV extinction is important to  understand the physical properties of dust.

\emph{GALEX} as the largest UV survey, its near-UV channel (NUV) covers the 2175 \,{\AA} bump, possibly a probe in polycyclic aromatic hydrocarbons \citep[PAHs;][]{2010ApJ...712L..16S}, and its far-UV channel (FUV) probes the steep far-UV rise of the extinction curve, associated with very small silicate grains \citep[WD01]{2001ApJ...548..296W}. Small grains dominate the heating of the diffuse ISM \citep{2003ApJ...587..278W} and may act as tracers of variation in grain size distribution as a whole. The \emph{GALEX} dataset provides us a chance to study the UV extinction in details.

Previously, we \citep{2018ApJ...861..153S} calculated the intrinsic colors and color excesses related to the \emph{GALEX} bands for about 25,000 A- and F-type dwarf stars. In this work, we will extend to many more stars, i.e. to about one million stars most of which are at high Galactic latitudes, and consequently to build the large-area UV extinction map. Taking the \emph{Gaia} blue ($\rm G_{BP}$) and red ($\rm G_{RP}$) band as reference, the UV extinction is used to retrieve the properties of dust at high Galactic latitude. In comparison with the previous studies of extinction map by multi-band photometry such as  \citet{2018A&A...616A...8A}, \citet{2019MNRAS.483.4277C} and \citet{2019ApJ...887...93G}, this size of sample is moderate. The advantage of this work lies in the special band -- ultraviolet. In addition, this work derives the intrinsic colors from spectroscopy rather than photometry, which is more accurate and reliable than statistical method.  The paper is organized as following. Section \ref{DATA} describes the data to be used, followed by Section \ref{method} on how the intrinsic colors and color excesses are calculated. The final Section \ref{resultanddiscussion} presents the results of the extinction map and discussions on the color excess ratios.

\section{Data Preparation}\label{DATA}

The basic idea to calculate the color excess complies with that adopted in \citet{2018ApJ...861..153S}. Firstly, the relation of intrinsic colors with stellar effective temperature and metallicity is derived for the given luminosity class. Here only dwarf stars are taken because giants are red and rarely detected in the \emph{GALEX} UV bands. In addition to the UV bands, the optical bands are needed in order to investigate the extinction law in the way of color excess ratio, and optical bands also provide an indicator to compare with other works. For its high accuracy and complete coverage of the sky, the selected optical survey is the \emph{Gaia} photometry. The stellar parameters are taken from the spectroscopic surveys -- LAMOST and GALAH.

\subsection{Photometric data from the GALEX and Gaia survey}

Unlike the spectroscopic surveys such as \emph{IUE}, \emph{FUSE} and \emph{SWIFT}, the \emph{GALEX} GR/6+7 \citep{2014Ap&SS.354..103B}\footnote{\url{http://dolomiti.pha.jhu.edu/uvsky}} provides  the largest high-quality photometric database in the ultraviolet. The two filters of \emph{GALEX} are \emph{FUV} ($\lambda_{{\rm eff}}=$1528 {\AA}, 1344-1786 {\AA}) and \emph{NUV} ($\lambda_{{\rm eff}}= $2310 {\AA}, 1771-2831 {\AA}). The catalog we use is GUVcat\_AIS\_fov055, with a total number of unique AIS sources (eliminating duplicate measurements) being 82,992,086, and a typical depth of \emph{FUV}=19.9 and \emph{NUV}= 20.8 AB mag \citep{2017ApJS..230...24B}.

In addition to astrometric information of proper motion and parallax, \emph{Gaia}/EDR3 \citep{2020arXiv201201533G} \footnote{\url{http://cosmos.esa.int/web/gaia}} provides photometry in three optical bands: the white-light G-band (330-1050 nm), the blue ($\rm G_{BP}$) and red ($\rm G_{RP}$) band by the prism photometers that collect low resolution spectrophotometric measurements of source spectral energy distributions over the wavelength ranges 330-680 nm and 630-1050 nm, respectively \citep{2020arXiv201201533G}.  There are 1.8 billion sources in the catalog, with the magnitude range from about 8 to 18 mag while G-band to 20.7 \citep{2016A&A...595A...2G}.

\subsection{Spectroscopic Data  from the LAMOST and GALAH survey}
The primary spectroscopic databaese we use comes from the LAMOST survey \citep{2015RAA....15.1095L} \footnote{\url{http://dr7.lamost.org/v1.1/}} which is dedicated to the objects over the entire available northern sky. The observable areas are mainly from -10$\degr$ to +90$\degr$ in declination. Taking 4000 spectra in a single exposure, the limiting magnitude in the $r$ band is as faint as 19 mag at a resolution R = 1800
\citep{2012RAA....12..723Z}. We use the LAMOST/DR7 catalog, which contains a total of 6,199,917 stars with the derived effective temperature $\Teff$,  surface gravity $\log g$, metal abundance $Z$ and their uncertainties.

In order to extend the sky coverage, we include the Galactic Archaeology with HERMES (GALAH) \citep{2020arXiv201102505B} \footnote{\url{http://galah-survey.org}} survey as a supplement in the southern hemisphere. Using the HERMES spectrograph at the Anglo-Australian Telescope, they collect high-resolution (R$\sim$28,000) spectra for 360 stars simultaneously. These stars are selected simply in magnitude (12$<V<$14), Galactic latitude ($|b|>10^{\circ}$) and declination (dec$<+10^{\circ}$). There are 588,571 unique stars in DR3. Observed between November 2013 and February 2019, they have derived radial velocities, stellar parameters and abundances.

\subsection{Combination of photometric and spectroscopic data}
The LAMOST/DR6 or GALAH/DR3 catalog is cross-matched separately with the \emph{Gaia}/EDR3 by a radius of 1$''$ first, which result in the LAMOST-\emph{Gaia} ('LG' hereafter, 6,142,479 stars) and the GALAH-\emph{Gaia} ('GG' hereafter, 588,553 stars) catalog for a large sky coverage in optical.

The LG or GG catalog is further cross-identified with the \emph{GALEX} GR6/7 catalog by a radius of 3$''$. Finally, there are two catalogs: LAMOST-\emph{Gaia}-\emph{GALEX} ('LGG' hereafter, 1,985,278 stars) and GALAH-\emph{Gaia}-\emph{GALEX} ('GGG' hereafter, 279,604 stars) which will be used to study the color excesses in the UV bands.

\section{Determination of the intrinsic colors in the \emph{GALEX} UV bands}\label{method}

\subsection{Source selection criteria}

\begin{figure}
\centering
\centerline{\includegraphics[scale=0.8]{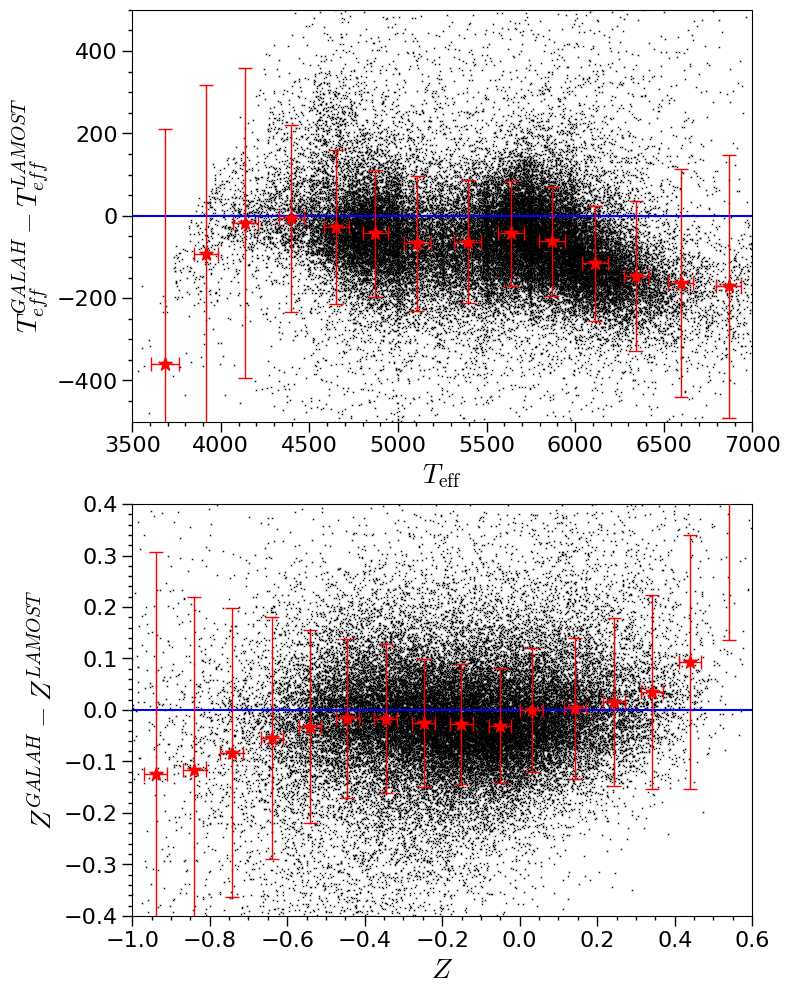}}
\caption{Comparison of $\Teff$ and $Z$ of the GALAH sample with the LAMOST. The red asterisks and bars denote the median values and their standard deviations.
\label{fig1}}
\end{figure}

The same as performed in \citet{2018ApJ...861..153S}, we limit the accuracy of $\Teff$ from the LAMOST and GALAH spectroscopic survey to be $\sigma_{\Teff}/\Teff \leq 5\%$. Only  the dwarf stars are selected by requiring $\log g \geq 4$ for $\Teff < 6500$\,K or $\log g \geq 3.5$ for $\Teff \geq 6500$\,K for LAMOST, and $\log g \geq 4$ for $\Teff < 5500$\,K or $\log g \geq 3.5$ for $\Teff \geq 5500$\,K for GALAH. The LG and GG catalogs are cross-matched and the stellar parameters of GALAH are compared with LAMOST. As illustrated in Figure~\ref{fig1}, the $\Teff$ of GALAH is lower at $\Teff > 6000$\,K, which agrees with the result of \citet{2019A&A...624A..19B}. So only the $\Teff$ within 4500\,K- 6000\,K is reliable, and fortunately most of the GALAH sources are covered. The metallicity is limited from -1.0 to 0.6 for LAMOST and -0.6 to 0.4 for GALAH (cf. Figure~\ref{fig1}), for both of which the error is limited to 0.3 dex. In addition, the accuracy of optical photometry is required to be better than 0.02 mag in the \emph{Gaia/}$\rm G_{BP}$ and $\rm G_{RP}$ band.

For the UV bands, the accuracy of photometry is limited to be better than 0.30 mag in the \emph{NUV} band and no restriction is set on the \emph{FUV} band because of the significantly fewer sources. Besides, the brightness is limited to \emph{NUV}$>$13.85 and \emph{FUV}$>$13.73 because high CTRs from UV-bright sources cause nonlinearity in the response, or saturation, due to the detector's dead-time correction being overtaken by the photon arrival rate for brighter sources \citep{2017ApJS..230...24B}. The stellar effective temperature is limited to  $\geq 5000$K for the \emph{NUV} band and  $\geq 6500$K for the \emph{FUV} band to reduce the influence of chromospheric activity at low  $\Teff$.  The metallicity complies with the rules in the optical for the \emph{NUV} band, but changed to -0.6 to 0.4 for the \emph{FUV} band, and its error is required to be smaller than 0.3 dex.

Consequently, there are 4,192,015 stars (3,980,392 stars in LG and 211,623 stars in GG) with the optical measurements and 1,271,780 stars with the \emph{NUV} photometry (1,156,319 stars in LGG and 115,461 stars in GGG)  as well as 59,247 with the \emph{FUV} photometry in LGG.

\subsection{Selection of the zero-extinction sources in the $C_{\rm G_{BP},G_{RP}}$  versus $\Teff$ and $Z$  Diagram}

The blue-edge method is a convenient and accurate way to obtain the relations of the intrinsic color index with the effective temperature. This method has been used by several works, e.g. \citet{2014ApJ...788L..12W}, \citet{2016ApJS..224...23X}, \citet{2017AJ....153....5J}, \citet{2018ApJ...855...12Z} and \citet{2020ApJ...891..137Z}. The essential idea is that the observed bluest color stands for the intrinsic one for a given luminosity class and effective temperature \citep{2001ApJ...558..309D}.

\begin{figure}
\centerline{\includegraphics[scale=0.6]{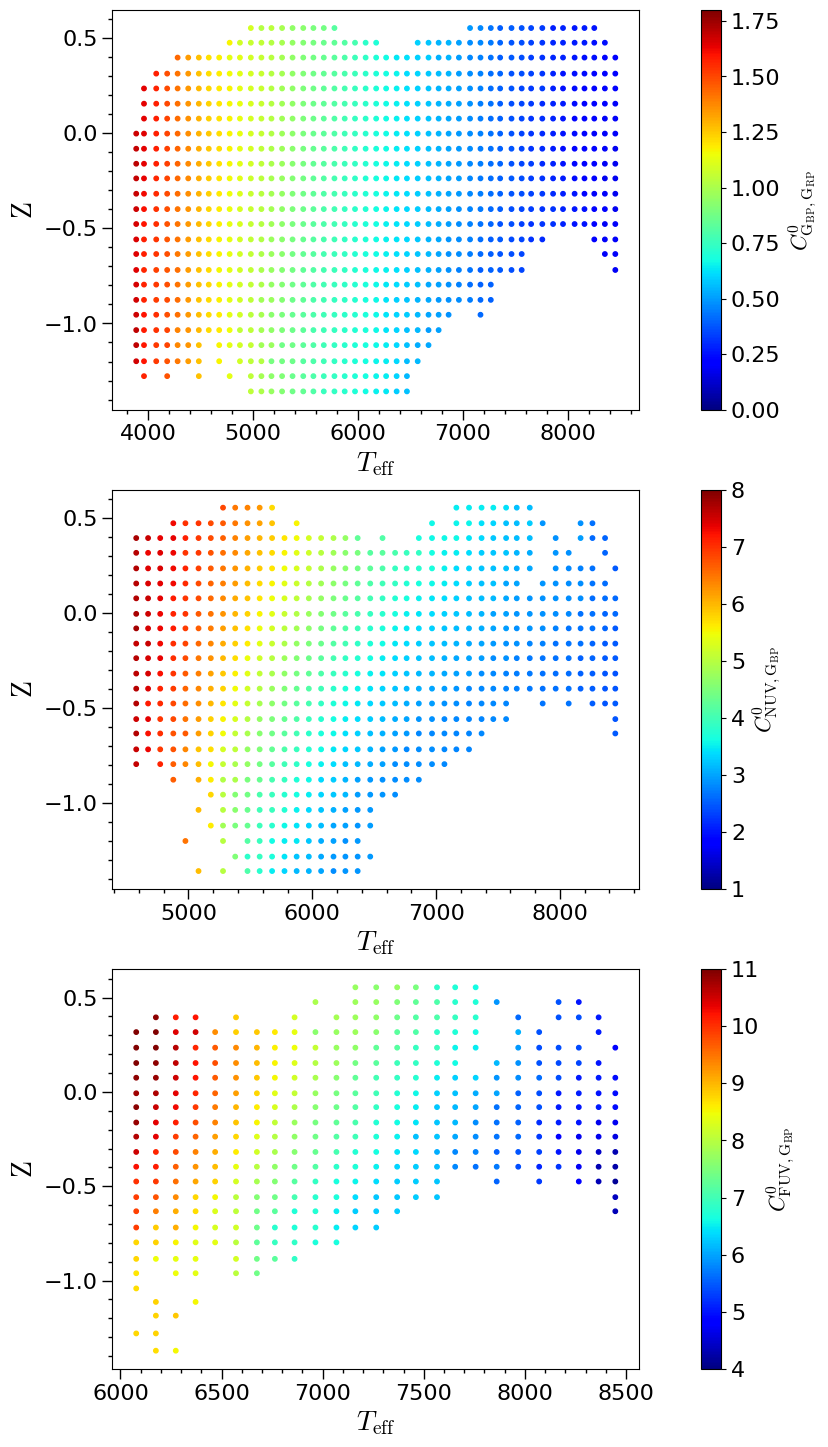}}
\caption{The intrinsic color index at the given $\Teff$ and $Z$ bin for the dwarf stars in the LAMOST/DR7 catalog. This is used to determine the  intrinsic color index $C^{0}_{\rm {G_{BP},G_{RP}}}$ (top), $C^{0}_{\rm {NUV,G_{BP}}}$ (middle) and $C^{0}_{\rm {FUV,G_{BP}}}$ (bottom) for a given $\Teff$ and $Z$.
\label{fig2}}
\end{figure}

For the UV bands, the influence of metallicity ($Z$) on the intrinsic color index is much more significant than in the optical or infrared and should be taken into account. Previously, \citet{2018ApJ...861..153S} chose the relatively high temperature sources to study the color excess ratio relevant to the UV band, where the effect of $Z$ is not serious and the stars are roughly divided into three groups. However, the present work handles the stars with a much large range of stellar parameters and the extinction at high latitudes is usually small,  the effect of $Z$ should be processed more delicately. For this purpose, we first divide the optical LG and GG sample  into groups with a step of 0.08-dex and 100-K in $Z$ and $\Teff$ respectively. Then the intrinsic color index for each group is obtained by the blue-edge method for $C_{\rm G_{BP},G_{RP}}$, i.e. the median color index of the bluest 5\% stars in the group chosen by a cyclic 3$\sigma$ elimination. In this way, there are many more (25 for $NUV$ and 13 for $FUV$) groups in metallicity. On the other hand, some groups have a small number of sources and the blue-edge method becomes unreliable. Such outliers are removed by finding the point deviating more than 1$\sigma$ from the mean of their neighbors, and then replaced by the mean of the neighbouring  3 $\times$ 3 block. The result is shown in the upper panel of Figure~\ref{fig2} where the range of $\Teff$ and $Z$ is slightly expanded in order to suppress the edge effect.

\subsection{Calculation of $C^{0}_{\rm UV,G_{BP}}$ with $\Teff$ and $Z$}

It should be noted that there is no clear cut at $\Teff$ for chromospheric activity. Even after limiting $\Teff \geq 5000$\,K and $\Teff \geq 6500$\,K for $NUV$ and $FUV$ respectively, the chromospheric activity at the low temperature still have a chance to bring about a bluer UV color. Therefore, the zero-extinction sources selected above from the $C_{\rm G_{BP},G_{RP}}$ vs. $\Teff$ and $Z$ diagram are used to calculate  $C^{0}_{\rm NUV,G_{BP}}$ and $C^{0}_{\rm FUV,G_{BP}}$ in each bin of $\Teff$ and $Z$, and the results of $C^{0}_{\rm NUV,G_{BP}}$ and $C^{0}_{\rm FUV,G_{BP}}$ are  shown respectively in the middle and bottom panels of Figure~\ref{fig2}.

It should be noted that the GALAH sample is one order of magnitude smaller than the LAMOST, thus $C^{0}_{\rm FUV,G_{BP}}$ is not calculated for them, and subsequently, the FUV-related extinction is neither obtained for the GALAH sample.

We find that the relation between the intrinsic color index and the stellar parameters is so complex that it is difficult to describe all the characteristics of the numerical solutions with one analytic formula. Therefore, a linear interpolation is performed between the lattice points to obtain the intrinsic color index of an individual star.

\begin{figure}
\centerline{\includegraphics[scale=0.6]{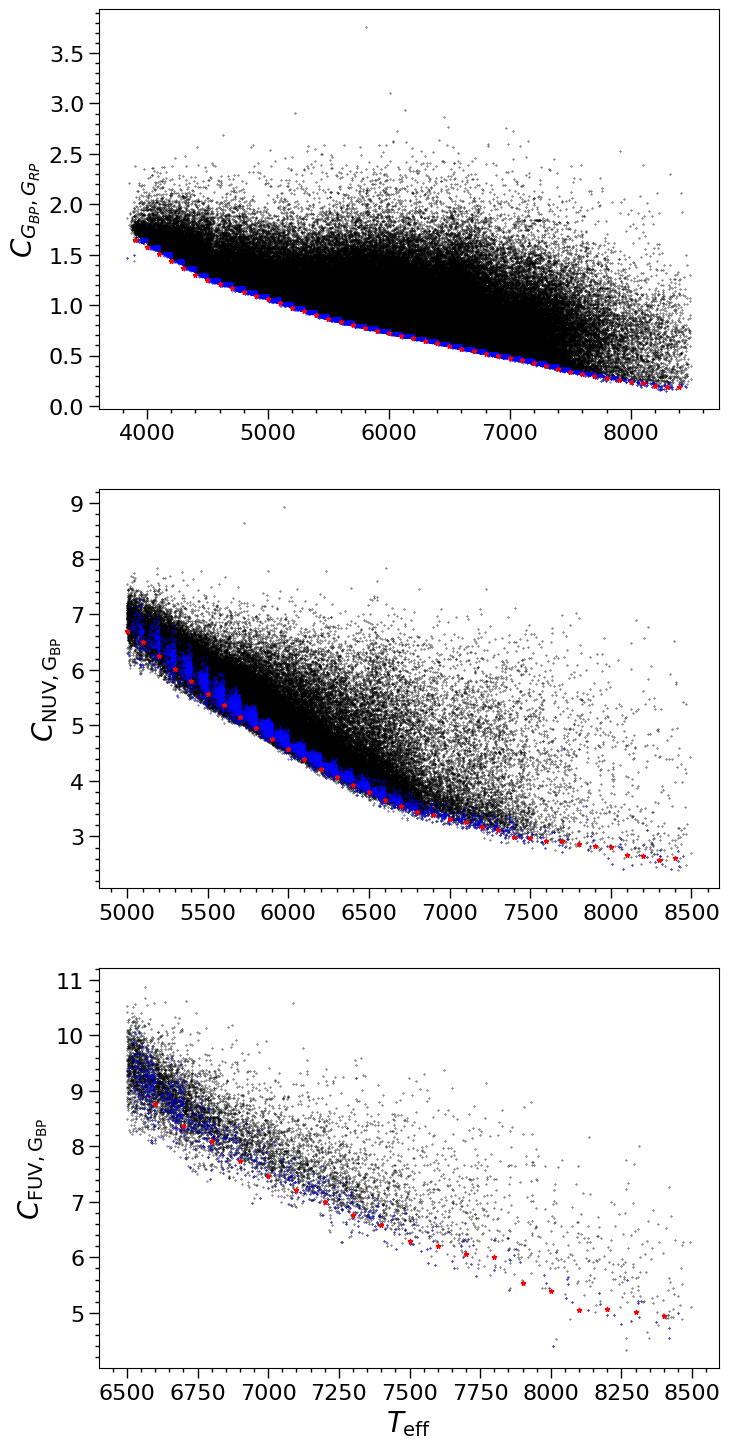}}
\caption{The change of the color index $C_{\rm {G_{BP},G_{RP}}}$ (top), $C_{\rm {NUV,G_{BP}}}$ (middle) and $C_{\rm {FUV,G_{BP}}}$ (bottom)  with $\Teff$ for the dwarf stars in the LAMOST/DR7 catalog with $-0.04 \leq Z <0.04 $. The black dots denote all the stars, the blue dots denote the selected zero-extinction sources and the red dots the determined intrinsic color index in the corresponding bin.
\label{fig3}}
\end{figure}

Figure~\ref{fig3} is an example displaying the change of the color index with $Z$ in the bin of [-0.04, 0.04]. It can be seen that the zero-extinction sources selected from optical is well concentrated in the blue edge of the  $C_{\rm UV,G_{BP}}$ vs. $\Teff$ and $Z$ diagrams.

\subsection{The error of color excesses}

The color excesses are straightforwardly calculated from the derived intrinsic and the observed colors, thus the errors of color excess can be estimated from that of the intrinsic and  observed colors. Since the majority of the sample comes from the LAMOST survey,  we take its error as the representative. The dispersion of the color index of the zero-extinction sources, i.e. the bluest 5\% stars in $C_{\rm G_{BP},G_{RP}}$ within each bin of $\Teff$ and $Z$ , is taken as the uncertainty of the intrinsic color. The dispersion is mainly caused by the error of stellar parameters and observed color index as well as the small change of intrinsic color index in a bin. Actually, the intrinsic color index obtained by linear interpolation should have smaller uncertainty than this dispersion. The median value of this dispersion is determined to be 0.007, 0.092 and 0.344 mag for $C^{0}_{\rm G_{BP}, G_{RP}}$, $C^{0}_{\rm NUV,G_{BP}}$ and $C^{0}_{\rm FUV,G_{BP}}$ respectively. This error originates mainly from stellar parameters and photometry. The error of observed colors comes from photometry whose median is  0.005, 0.071 and 0.270 mag for $C_{\rm G_{BP}, G_{RP}}$, $C_{\rm NUV,G_{BP}}$ and $C_{\rm FUV,G_{BP}}$ respectively. The combined error median becomes 0.009, 0.128 and 0.454 mag respectively. The error is calculated for each source individually in this way.

\begin{figure}
\centering
\centerline{\includegraphics[scale=0.8]{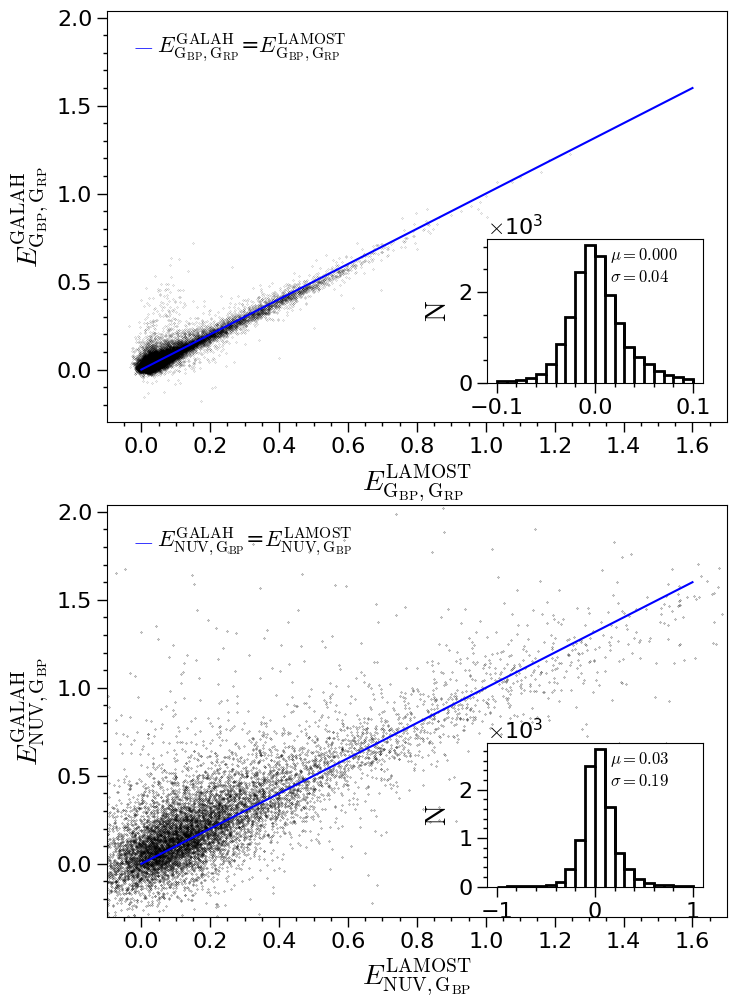}}
\caption{Comparison of $E_{{\rm G_{BP},G_{RP}}}$ and $E_{\rm NUV,G_{BP}}$ of the LAMOST sample with the GALAH. The inset shows the mean and standard deviation of the difference.
\label{fig4}}
\end{figure}

Another practical way to estimate the error of the color excess is to compare the identical sources of the LAMOST and GALAH sample, which contains 17,472 common stars in $E_{{\rm G_{BP},G_{RP}}}$ and 9,870 stars in $E_{\rm NUV,G_{BP}}$. The result is displayed in Figure~\ref{fig4}, where the systematic difference should be small and is neglected. The dispersion is 0.04 mag in $E_{\rm G_{BP}, G_{RP}}$ and 0.19 mag in $C_{\rm NUV,G_{BP}}$, which is comparable to the error derived above from the uncertainty of intrinsic and observed colors. For $E_{\rm FUV,G_{BP}}$, we specifically calculated the color excess of GALAH in the $\Teff$ range of 6500-7000 K, and compared it with that of LAMOST. From 333 common sources, the dispersion is about 0.40, which is also consistent with the estimated uncertainty.

\subsection{Comparison the optical color excess with the Green+2019 work}

\begin{figure}
\centering
\centerline{\includegraphics[scale=0.8]{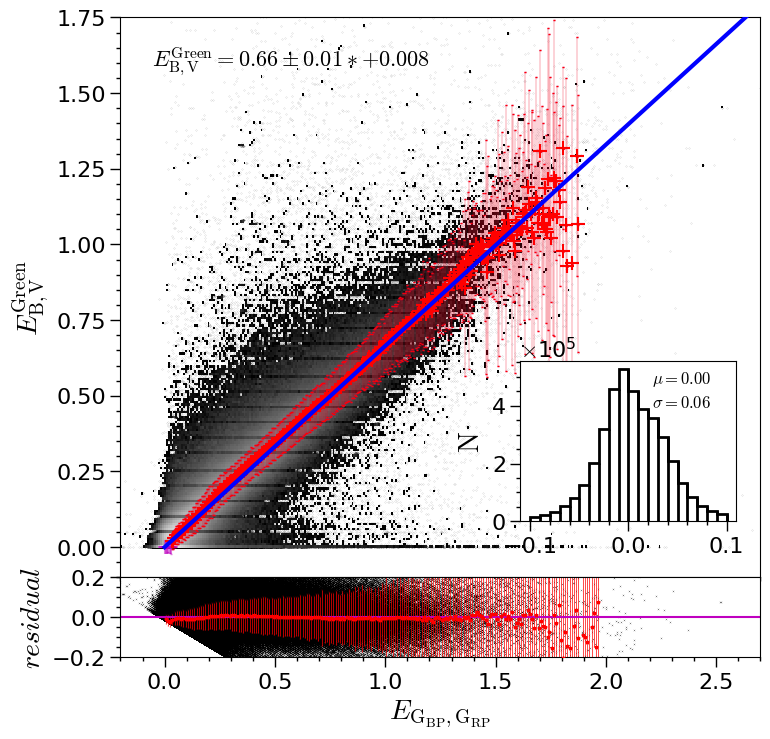}}
\caption{Comparison of $E_{{\rm G_{BP},G_{RP}}}$ derived in this work with $ E_{\rm B,V}^{\rm Green}$ from \citet{2019ApJ...887...93G}. The top panel displays the linear relation between them with the mean and the standard deviation of the residuals displayed in the inset, while the bottom panel shows the residuals change with $E_{{\rm G_{BP},G_{RP}}}$.
\label{fig5}}
\end{figure}

As a by-product, the $E_{{\rm G_{BP},G_{RP}}}$ value is obtained  for 4,169,769 sources composed of 3,959,547 stars in LG and 210,222 stars in GG. For the given coordinate and distance \footnote{The distance here is calculated from \emph{Gaia}-measured parallax and its error by Smith-Eichhorn correction method \citep{1996MNRAS.281..211S,2018A&A...616A...9L}} from \emph{Gaia} of a star, the $ E_{\rm B,V}^{\rm Green}$ value by \citet{2019ApJ...887...93G} can be retrieved by the python code dustmap\footnote{\url{https://dustmaps.readthedocs.io/}}. The linear fitting (see Section \ref{ratio}) between these two values in the upper panel of Figure~\ref{fig5} yields $E_{\rm B,V}^{\rm Green} = 0.66 \pm 0.01 * E_{{\rm G_{BP},G_{RP}}} -0.008$, with the residual of a mean of zero and a width of about 0.06\,mag whose distribution is displayed in the lower panel of Figure~\ref{fig5}, which is comparable to the 0.04 mag internal precision.  Basically there is no systematic difference with the Green+2019 work, and the standard deviation is small as well. Nevertheless, there are some stars for which $ E_{\rm B,V}^{\rm Green} =0$ while our derived $E_{{\rm G_{BP},G_{RP}}}$ span a range up to 2.0\,mag. There are 21 sources with $E_{\rm B,V}^{\rm Green} \leq 0.05$ mag while $E_{{\rm G_{BP},G_{RP}}} \geq 2$ mag, which are  $>3 \sigma$ outliers in the linear fitting. Examination of their spectroscopic and photometric parameter finds no apparent problem. These sources are checked in the SIMBAD\footnote{\url{http://simbad.u-strasbg.fr/simbad/}}  database, two of which are found to have high proper-motion, and we calculated the distance to be zero with the method of \citet{1996MNRAS.281..211S} and \citet{2018A&A...616A...9L}. Meanwhile,  no distance is available in the Gaia/DR2 or EDR3 catalog \citep{2018AJ....156...58B,2021AJ....161..147B}. For sure, the zero distance is incorrect possibly caused by not considering the high proper motion so that the color excess from \citet{2019ApJ...887...93G} can be not corresponding. But this is not the case for the other sources which may be dense but small dust cloud (e.g. Bok globules)  that cannot be resolved at the low-resolution map of interstellar extinction or dust emission by PLANCK. The details of these objects are listed in Table~\ref{table1} with the position, distance and extinction. As a whole, the residual (defined as $E_{\rm B,V}^{\rm Green} - (0.66*E_{{\rm G_{BP},G_{RP}}}+0.008)$) in the lower panel of Figure~\ref{fig5} shows no obvious trend with $E_{{\rm G_{BP},G_{RP}}}$, and it is generally very close to 0. The other result is that $E_{\rm B,V}^{\rm Green}/E_{{\rm G_{BP},G_{RP}}}=0.66 \pm 0.01$, which is smaller than $E_{{\rm B,V}} / E_{{\rm G_{BP},G_{RP}}}=0.71 \pm 0.01$ (see Section \ref{ratio}) and $E_{{\rm B,V}} / E_{{\rm G_{BP},G_{RP}}}=0.76$ by \citet{2019ApJ...877..116W}.


\section{Result and Discussion} \label{resultanddiscussion}

\subsection{The Ultraviolet Color Excess $E_{{\rm NUV,G_{BP}}}$ and $E_{{\rm FUV,G_{BP}}}$ }

With the determined intrinsic color index, the NUV color excess $E_{{\rm NUV,G_{BP}}}$ is calculated for 1,244,504 sources (1,130,412 sources from LGG and the 114,092 sources from GGG), and the FUV color excess $E_{{\rm FUV,G_{BP}}}$ is calculated for 56,123 sources from LGG. In comparison, \citet{2018ApJ...861..153S} calculated $E_{{\rm NUV,G_{BP}}}$ and $E_{{\rm FUV,G_{BP}}}$ for 25,496 and 4,255 A- and F-type stars respectively. The huge increase to a million stars is brought by a couple of improvements: (1) the extension to much lower effective temperature, from previous 6500\,K for NUV (7000\,K for FUV) to present 5000\,K (6500\,K)  , which is the major factor, (2) the updated LAMOST data release adding one million stars in their entire catalog, (3) less strict constraints on the photometric uncertainty in the UV bands, and (4) the inclusion of the GALAH database.

\begin{figure}
\centering
\centerline{\includegraphics[scale=0.4]{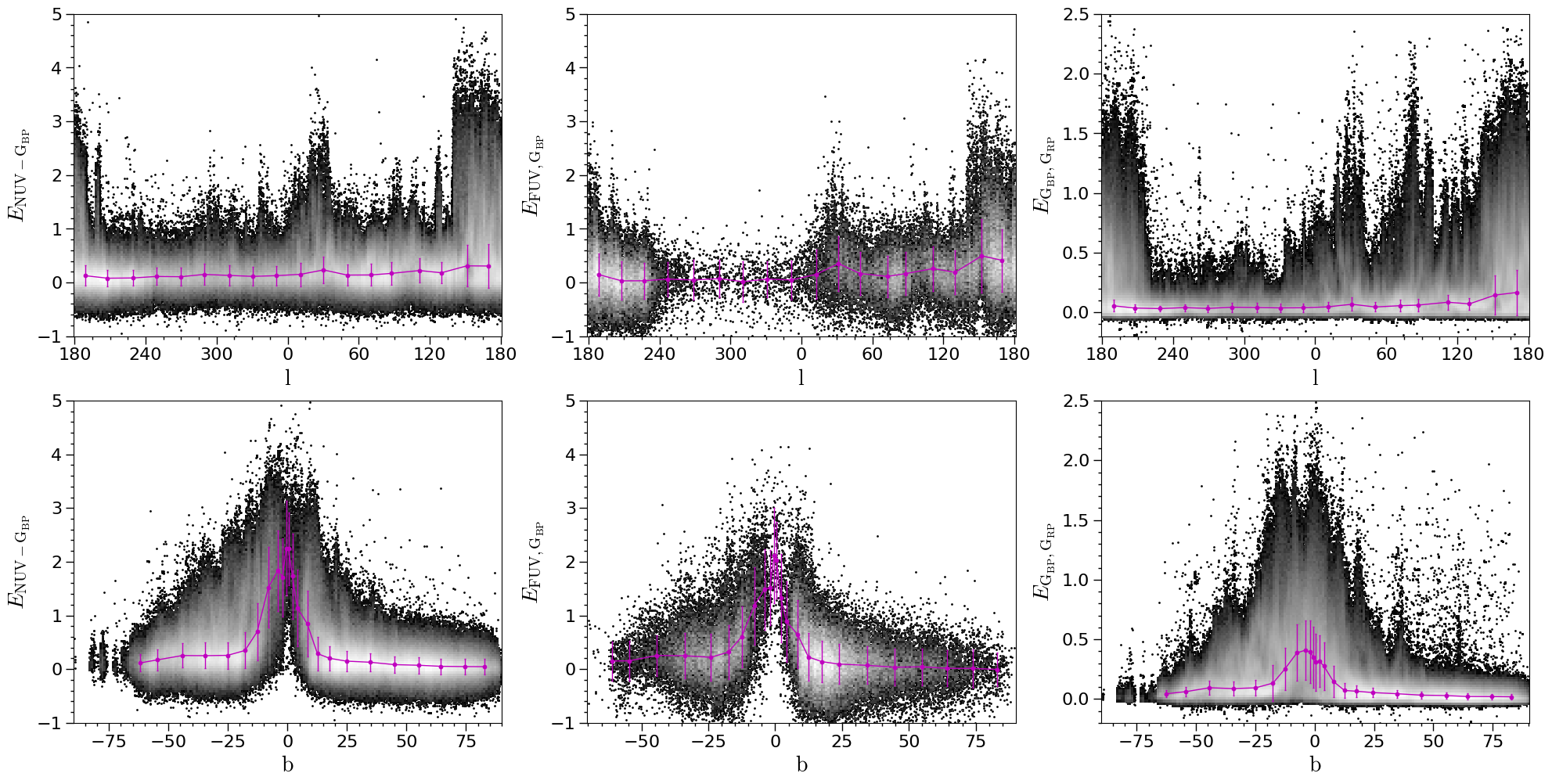}}
\caption{The change of $ E_{{\rm NUV,G_{BP}}}$ (left), $ E_{{\rm FUV,G_{BP}}}$ (middle) and $ E_{{\rm G_{BP},G_{RP}}}$ (right) with  Galactic longitude (top) and latitude (bottom). The magenta points and lines are the median values of background.
\label{fig6}}
\end{figure}

For the distribution of color excess, there are 207,942 and 16,822, i.e. 17\% sources with $E_{{\rm NUV,G_{BP}}} < 0$ and 30\% sources with $E_{{\rm FUV,G_{BP}}} < 0$ , which is much higher than the proportion (4\%) with $E_{{\rm G_{BP},G_{RP}}}<0$. In addition to the relatively large photometric error in the UV bands, the chromospheric activity can also lead to the negative color excess. Nevertheless, most stars show a positive $E_{{\rm NUV,G_{BP}}}$ and $E_{{\rm FUV,G_{BP}}}$ with the peak both around 0.07 mag, much more significant than the optical excess whose peak is around 0.03 mag.  The change with Galactic longitude and  latitude is displayed in Figure~\ref{fig6}, where the median value denoted by the magenta line is obtained by an iterative 3$\sigma$ clipping. Because the \emph{GALEX} observation is far from complete towards the Galactic plane, the variation along the longitude lacks a large-scale trend except that the Galactic center and anticenter directions have relatively large background color excess, otherwise dominated by some small-scale local structures from star-forming regions. On the other hand, the change with latitude is very evident that the color excess starts to increase rapidly with the decreasing Galactic latitude at $|b|<20^{\circ}$ and the maximum value reaches to 2.25 mag and 2.12 mag in $E_{{\rm NUV,G_{BP}}}$  and $E_{{\rm FUV,G_{BP}}}$ towards the Galactic plane.

\subsection{The Color Excess Ratio}

\subsubsection{$E_{{\rm NUV,G_{BP}}} / E_{{\rm G_{BP},G_{RP}}}$  and $E_{{\rm FUV,G_{BP}}} / E_{{\rm G_{BP},G_{RP}}}$ }\label{ratio}

\begin{figure}
\centering
\centerline{\includegraphics[scale=0.8]{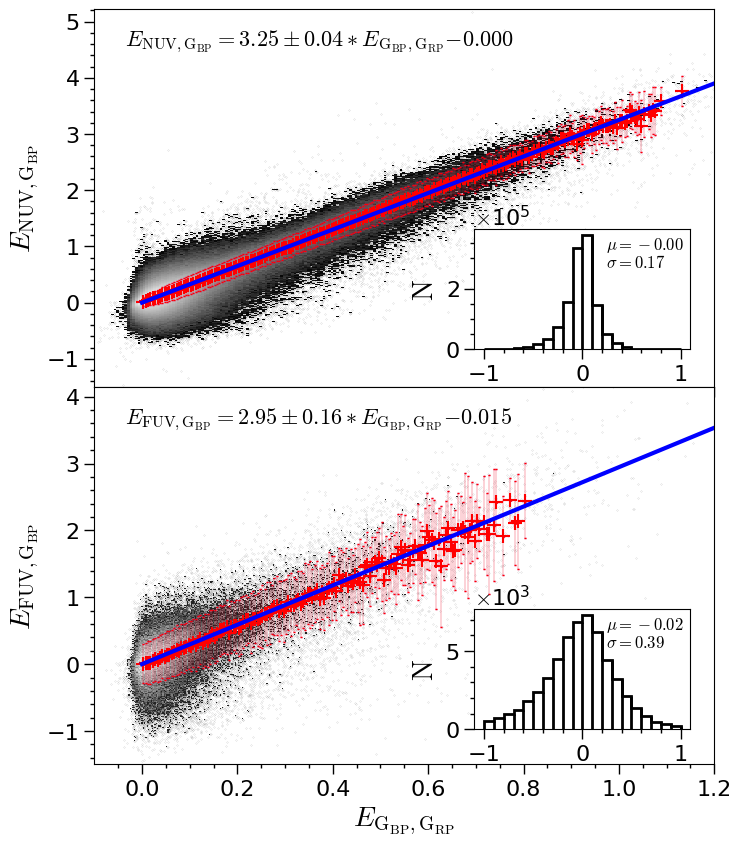}}
\caption{Linear fitting of the color excesses $ E_{{\rm NUV,G_{BP}}}$ to $ E_{{\rm G_{BP},G_{RP}}}$ (top) and $ E_{{\rm FUV,G_{BP}}}$ to $ E_{{\rm G_{BP},G_{RP}}}$ (bottom). The gray-scale encodes the source densities, while the red crosses and bars denote the median value and its standard deviation of each bin (see Section \ref{ratio} for details). The inset shows the distribution of the residuals with its median and standard deviation.
\label{fig7}}
\end{figure}

In order to suppress the influence of numerous low-extinction stars, the median values in a bin of 0.005 mag of $E_{{\rm G_{BP},G_{RP}}}$ are calculated with iteratively clipping stars beyond 3$\sigma$ of the median. In addition, a bin with less than 10 sources are not used in linear fitting.  As shown in Figure~\ref{fig7}a, the Markov Chain Monte Carlo (MCMC) procedure is performed \citep{2013PASP..125..306F} for the linear fitting. The best estimates are the median values (50th percentile) of the posterior distribution and the uncertainties are derived from the 16th and 84th percentile values, which result in a relation of $E_{{\rm NUV,G_{BP}}}$ and $E_{{\rm G_{BP},G_{RP}}}$  as $E_{{\rm NUV,G_{BP}}} = 3.25 \pm 0.04 * E_{{\rm G_{BP},G_{RP}}} - 0.0003$.

The number of sources in the far ultraviolet band is much fewer than in the near ultraviolet. The same method is used for linear fitting, which yields $E_{{\rm FUV,G_{BP}}} = 2.95 \pm 0.16 * E_{{\rm G_{BP},G_{RP}}} + 0.015$ shown in Figure~\ref{fig7}b.

As mentioned above, the linear fitting yields a color excess ratio of $E_{{\rm NUV,G_{BP}}} / E_{{\rm G_{BP},G_{RP}}} = 3.25 \pm 0.04$ and $E_{{\rm FUV,G_{BP}}} / E_{{\rm G_{BP},G_{RP}}} = 2.95 \pm 0.16$, which involving both the \emph{Gaia} and \emph{GALEX} filters has not been determined before. In order to check the consistency with previously determined $E_{{\rm NUV,B}} / E_{{\rm B,V}}$ and $E_{{\rm FUV,B}} / E_{{\rm B,V}}$, we further cross-identify the LG catalog with the APASS photometric survey in the $B$ and $V$ bands.  The same linear fitting method in previous subsection is used that yields $E_{{\rm B,G_{BP}}} / E_{{\rm G_{BP},{G_{RP}}}}=0.58 \pm 0.01$ and $E_{{\rm B,V}} / E_{{\rm G_{BP},G_{RP}}}=0.71 \pm 0.01$. With $ E_{{\rm NUV,G_{BP}}} / E_{{\rm G_{BP},G_{RP}}}$=3.25 and $ E_{{\rm FUV,G_{BP}}} / E_{{\rm G_{BP},G_{RP}}}$=2.95, the values of $E_{{\rm NUV,B}} / E_{{\rm B,V}}$  and $E_{{\rm FUV,B}} / E_{{\rm B,V}}$ are 3.76 $\pm$ 0.08 and 3.34 $\pm$ 0.23, which is very close to 3.77  $\pm$ 0.08 and 3.39 $\pm$ 0.17 by \citet{2018ApJ...861..153S}. In addition, the optical color excess ratio can be compared to the results of \citet{2019ApJ...877..116W} who used red clump stars to determine $E_{{\rm B,G_{BP}}} / E_{{\rm G_{BP},G_{RP}}}=0.72$ and $E_{{\rm B,V}} / E_{{\rm G_{BP},G_{RP}}}=0.76$. This agrees with ours at $E_{{\rm B,V}} / E_{{\rm G_{BP},G_{RP}}}$, but has a higher $E_{{\rm B,G_{BP}}} / E_{{\rm G_{BP},G_{RP}}}$. This may be explained by using different tracers. This work takes dwarfs as tracers, while \citet{2019ApJ...877..116W} uses red clump stars. Definitely, the dwarfs here brings about a shorter effective wavelength of the wide filter $G_{BP}$ than the red clump stars.

\subsubsection{The color excess ratio $E_{\rm FUV,NUV}$/$E_{\rm G_{BP},G_{RP}}$}

\begin{figure}
\centering
\centerline{\includegraphics[scale=0.8]{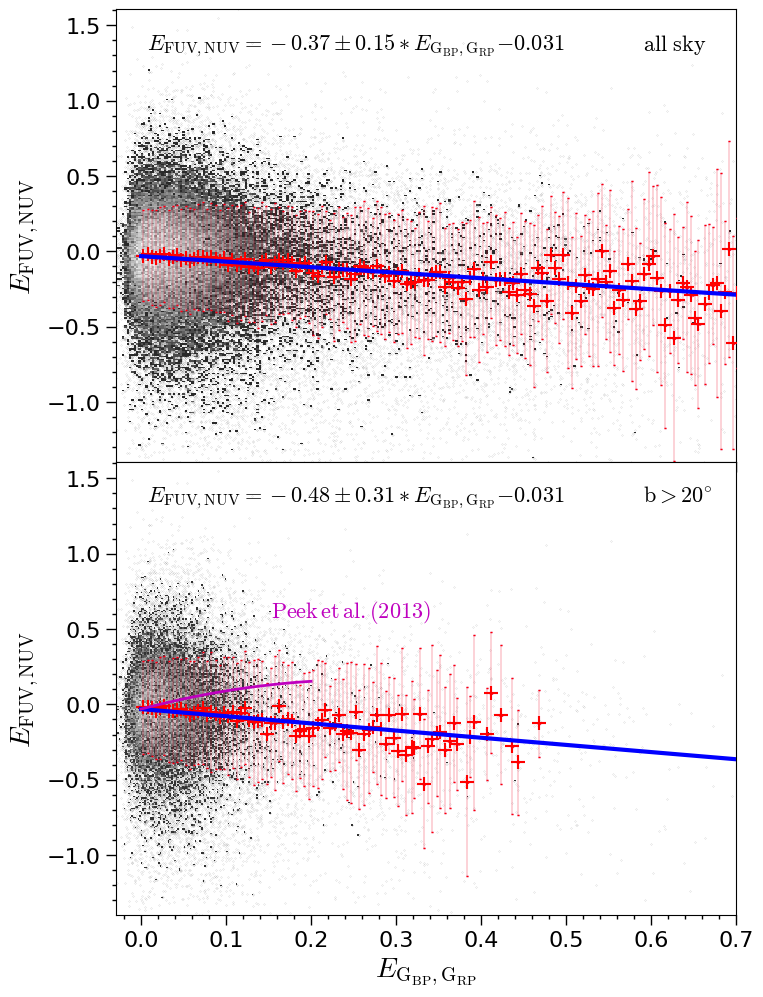}}
\caption{The same as Figure \ref{fig7}, but for the relations of the color excesses, $E_{{\rm FUV,NUV}}$ and $ E_{{\rm G_{BP},G_{RP}}}$ of all the LGG sample stars (top) as well as $ E_{{\rm FUV,NUV}}$ and $ E_{\rm B,V}^{SFD}$  of the LGG sample stars at $|b|>20^{\circ}$ (bottom) where the results of \citet{2013ApJ...771...68P} are shown for reference.
\label{fig8}}
\end{figure}

The linear fitting in Figure~\ref{fig8}a yields the relation between $E_{\rm FUV,NUV}$ and $E_{\rm G_{BP},G_{RP}}$, i.e. $E_{\rm FUV,NUV}= (-0.37 \pm 0.15) *E_{\rm G_{BP},G_{RP}}-0.031$, which agrees with $-0.30 \pm 0.16$ calculated in Section \ref{ratio} from $E_{{\rm FUV,G_{BP}}} / E_{{\rm G_{BP},G_{RP}}} - E_{{\rm NUV,G_{BP}}} / E_{{\rm G_{BP},G_{RP}}}$. Again, a negative value of $E_{\rm FUV,NUV}$ is found on average which confirms the result of \citet{2018ApJ...861..153S}. This is particularly true for  the relatively large and thus more reliable  color excesses. This result can be understood because the effective wavelength of the $NUV$ filter is around the peak of the 2175${\AA}$ bump for the sample stars while that of the $FUV$ filter is on the wing of the hump of the extinction curve.

Using the galaxies at high latitudes as tracers, \citet{2013ApJ...771...68P} studied the color excess $E_{\rm FUV,NUV}$ of \emph{GALEX} and found it to be positive, i.e. the extinction in the $NUV$ band is smaller than in the $FUV$ band. On the contrary, \citet{2018ApJ...861..153S} found $E_{\rm FUV,NUV}$ to be negative for about 4000 stars which distribute in the whole \emph{GALEX} area. Though this work uses a very similar method as \citet{2018ApJ...861..153S}, the catalogs are updated and a much larger sample is useful, which makes it feasible to study the $FUV$ extinction at high latitude alone.

The stars with $|b|>20^{\circ}$ are further picked up to see if the color excess $E_{\rm FUV,NUV}$ becomes positive at high-latitude as found by \citet{2013ApJ...771...68P}. The linear fitting yields $E_{\rm FUV,NUV}= (-0.48 \pm 0.31)*E_{\rm G_{BP},G_{RP}}-0.031$ shown in Figure~\ref{fig8}b. Basically, the UV-to-optical color excess ratio at high-latitude shows no difference with the entire sample.  Nevertheless, at both high and low latitude, some fraction of stars show positive $E_{\rm FUV,NUV}$ at $E_{\rm G_{BP},G_{RP}}<0.1$  which is the main range of \citet{2013ApJ...771...68P}. This may lead them to conclude a positive $E_{\rm FUV,NUV}$. In our sample, this range of color excess seems to be greatly affected by the errors, and a positive $E_{\rm FUV,NUV}$ is inconsistent with the general trend.

\subsection{Variation of the color excess ratios}

\subsubsection{Variation with the Galactic latitude and $E_{\rm G_{BP},G_{RP}}$}\label{selection}

The color excess is an indicator of the interstellar extinction law, so its variation may reflect the change of the dust properties.

\begin{figure}
\centering
\centerline{\includegraphics[scale=0.6]{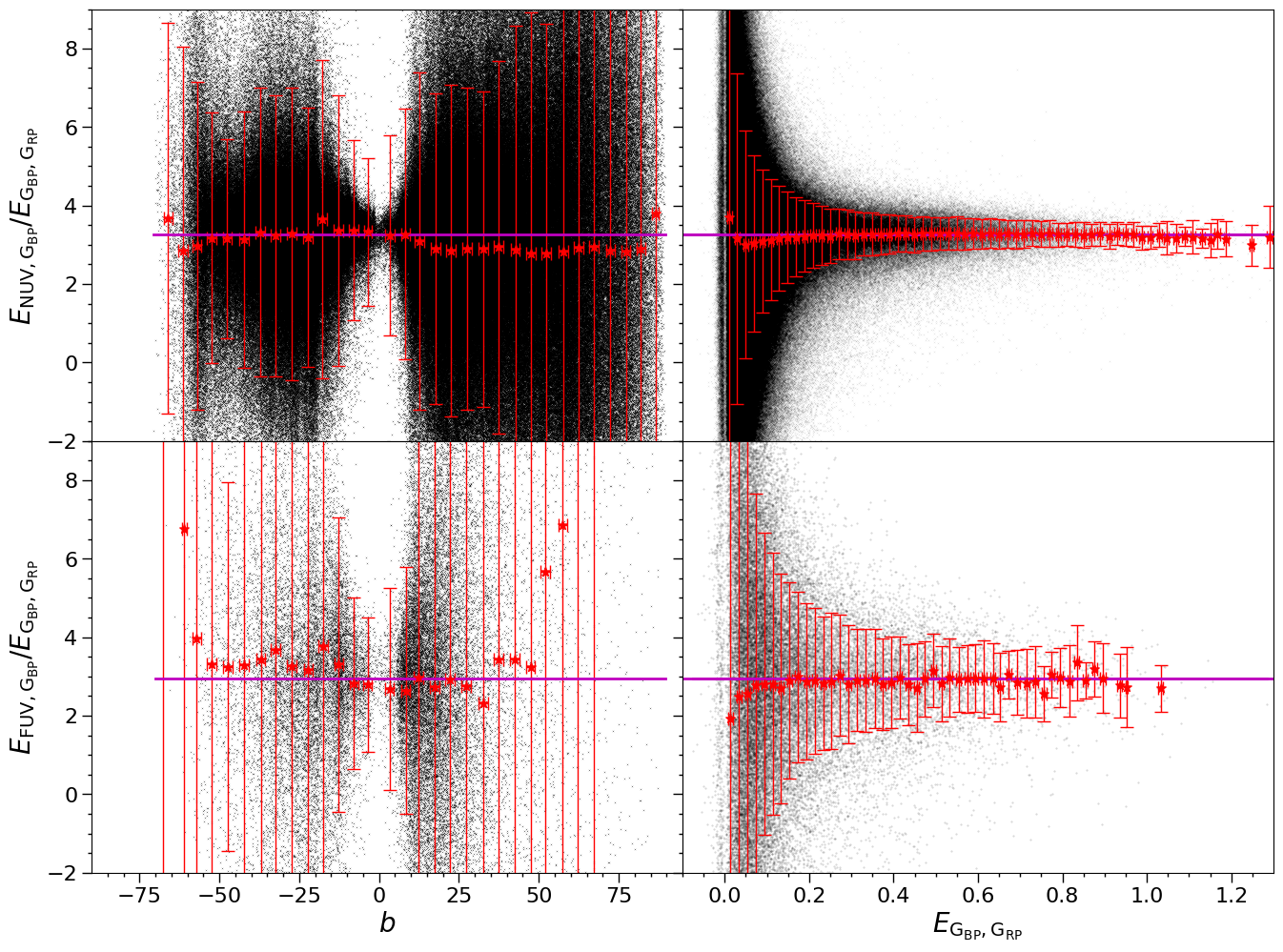}}
\caption{The change of color excess ratio with the Galactic latitude (left panel) and $E_{\rm G_{BP},G_{RP}}$ (right panel) for $ E_{\rm NUV,G_{BP}}$/$E_{\rm G_{BP},G_{RP}}$ (upper) and $ E_{\rm FUV,G_{BP}}$/$E_{\rm G_{BP},G_{RP}}$ (lower). The black dots denote the individual source, while the red asterisks and bars denote the ratio of the median value and its standard deviation of the color excesses in a bin of $5^{\circ}$ in the Galactic latitude (left panel) or in a bin of 0.02 mag in $E_{\rm G_{BP},G_{RP}}$ (right).
\label{fig9}}
\end{figure}

The variation of the color excess ratios is investigated by using only the LGG sources in order  to avoid the systematic difference in the stellar parameters between the LAMOST and GALAH survey. The change of $E_{\rm NUV,G_{BP}}/E_{\rm G_{BP},G_{RP}}$ (upper panel) and  $E_{\rm FUV,G_{BP}}/E_{\rm G_{BP},G_{RP}}$ (lower panel) with the Galactic latitude is displayed in the left panels of Figure~\ref{fig9}, where the color excess ratio of each star is denoted by black dots, the ratio of the median color excesses of the stars in a bin of $5^{\circ}$ in the Galactic latitude is denoted by red asterisks and the standard deviation of each bin is denoted by the red bars. The number of sources is required to be more than 10 for a bin to be considerable. Generally speaking, the color excess ratios have large dispersions. Within the dispersion, both $E_{\rm NUV,G_{BP}}/E_{\rm G_{BP},G_{RP}}$ and  $E_{\rm FUV,G_{BP}}/E_{\rm G_{BP},G_{RP}}$ show no significant variation with the latitude. The ratio $E_{\rm NUV,G_{BP}}/E_{\rm G_{BP},G_{RP}}$ is very constant along the latitude, while there is a small increase at the high latitude but only in a single bin. Meanwhile, $E_{\rm FUV,G_{BP}}/E_{\rm G_{BP},G_{RP}}$ seems increasing at the high latitude, but this tendency may be unreliable because the small number of sources brings about large uncertainty.

Similarly, the variation of the color excess ratios with the color excess $E_{\rm G_{BP},G_{RP}}$ is displayed in the right panels of Figure~\ref{fig9} for  $E_{\rm NUV,G_{BP}}/E_{\rm G_{BP},G_{RP}}$ (upper panel) and $E_{\rm FUV,G_{BP}}/E_{\rm G_{BP},G_{RP}}$ (lower panel) respectively, where the bin-size is 0.02 mag in $E_{\rm G_{BP},G_{RP}}$ and the number of sources in each bin also should be more than 10. Consistent with the case of the Galactic latitude, there is no significant variation with the color excess. Nevertheless, $E_{\rm NUV,G_{BP}}/E_{\rm G_{BP},G_{RP}}$ does increase a little at $E_{\rm G_{BP},G_{RP}} < 0.1$, which agrees with that it increases at the high latitude. On the other hand, $E_{\rm FUV,G_{BP}}/E_{\rm G_{BP},G_{RP}}$ decreases at  $E_{\rm G_{BP},G_{RP}} < 0.1$, which contradicts with that it probably increases at the high latitude.

\subsubsection{Effects of stellar effective temperature and interstellar extinction on the color excess ratios}

Before discussing the implication of the variation of the color excess ratios on the dust properties, the influence of stellar effective temperature and interstellar extinction should be clarified because they both would shift the effective wavelengths ($\lambda_{\rm eff}$) of the photometric filters and change the ratio, which is significant when the extinction is small like in this work. For this purpose, we choose four stars typical of the sample sources, i.e the dwarf stars with $\log g= 4 $, $Z=0 $,  and $\Teff = $5000 K, 6000 K, 7000 K and 8000 K respectively, whose spectral energy distribution is calculated by the ATLAS9 model. The range of $E_{\rm G_{BP},G_{RP}}$ is from 0.0 to 1.0. The SED is convolved with the F99 extinction law at $\RV = 3.1$ and the filter's response curve to calculate the effective wavelengths, the corresponding color excess and the color excess ratio. The results are illustrated in  Figure~\ref{fig10}. It can be seen that $E_{\rm NUV,G_{BP}}/E_{\rm G_{BP},G_{RP}}$ is systematically larger for high-$\Teff$ which is caused mainly by their larger shift to the shorter $\lambda_{\rm eff}$ of the $NUV$ band, and the difference between low-$\Teff $ and high-$\Teff$ diminishes with the color excess increasing. On the contrary, $E_{\rm FUV,G_{BP}}/E_{\rm G_{BP},G_{RP}}$ is larger for low-$\Teff$ and the difference with high-$\Teff$ is not reduced by increasing the color excess in the given range. This is because $\lambda_{\rm eff}^{\rm FUV}$  is almost constant in this $\Teff $ range and the change of the ratio originates mainly from the shift of $\lambda_{\rm eff}^{\rm G_{RP}}$.

\subsubsection{Change of $\Teff$ with the Galactic latitude and  $E_{\rm G_{BP},G_{RP}}$}

The change of $\Teff$ with the Galactic latitude and  $E_{\rm G_{BP},G_{RP}}$ is examined in order to peel off the influence of $\Teff$. Figure~\ref{fig11} shows that $\Teff$ changes with the Galactic latitude and $E_{\rm G_{BP},G_{RP}}$, where the upper panel takes into account the sources in the $NUV$ catalog and the lower panel the sources in the $FUV$ catalog. The average $\Teff$ in a bin of $5^{\circ}$ in the Galactic latitude (red asterisks with the red bar for the standard deviation) increases from about 5700 K in the high Galactic latitude region to about 6500 K in the low Galactic latitude region. Meanwhile, the average $\Teff$ in a bin of 0.02 mag in $E_{\rm G_{BP},G_{RP}}$ (red asterisks with the red bar for the standard deviation) increases from about 5800 K at the low extinction to about 7500 K at the high extinction. For the FUV catalog, the average $\Teff$ increases from about 6650 K in the high Galactic latitude region to 7600 K in the low Galactic latitude region, and the average $\Teff$ increases from about 6700 K at low extinction to about 8200 K at high extinction. In a word, $\Teff$ increases with the decreasing Galactic latitude and with the increasing extinction, and the steepness is stronger in the $FUV$ than in the $NUV$ band.

Considering the effects of $\Teff$ and $E_{\rm G_{BP},G_{RP}}$ on the color excess ratios in combination with the change of $\Teff$ with the Galactic latitude and  $E_{\rm G_{BP},G_{RP}}$, the following would be expected: (1) at the high latitude, $\Teff$ is low and $E_{\rm G_{BP},G_{RP}}$ is small, thus  $E_{\rm NUV,G_{BP}}/E_{\rm G_{BP},G_{RP}}$ should be smaller than the average and $E_{\rm FUV,G_{BP}}/E_{\rm G_{BP},G_{RP}}$ should be approximate to the average; (2) at the low latitude, $\Teff$ is high and $E_{\rm G_{BP},G_{RP}}$ is big, thus  $E_{\rm NUV,G_{BP}}/E_{\rm G_{BP},G_{RP}}$ should be bigger than the average and $E_{\rm FUV,G_{BP}}/E_{\rm G_{BP},G_{RP}}$ should also be approximate to the average; (3) at the small $E_{\rm G_{BP},G_{RP}}$, $\Teff$ is low, which the same as Case (1); (4) at the large $E_{\rm G_{BP},G_{RP}}$, $\Teff$ is high, which is the same as Case (2). Consequently, $E_{\rm NUV,G_{BP}}/E_{\rm G_{BP},G_{RP}}$ is expected to increase with increasing $E_{\rm G_{BP},G_{RP}}$ or decreasing latitude while $E_{\rm FUV,G_{BP}}/E_{\rm G_{BP},G_{RP}}$ is expected to be more or less constant independent of the latitude and color excess.

\begin{figure}
\centering
\centerline{\includegraphics[scale=0.8]{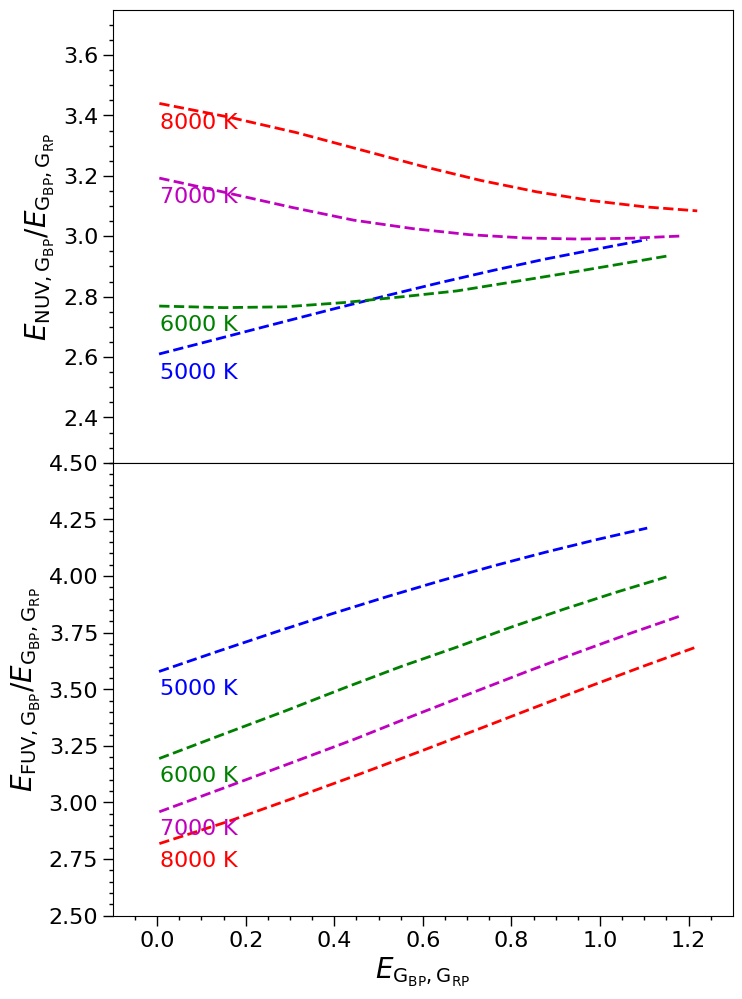}}
\caption{The change of the color excess ratio due to the shifts of the effective wavelength of the filters by the extinction and stellar $\Teff$ (see text for details) for $ E_{\rm NUV,G_{BP}}$/$E_{\rm G_{BP},G_{RP}}$ (upper panel) and  $ E_{\rm FUV,G_{BP}}$/$E_{\rm G_{BP},G_{RP}}$ (lower panel).
\label{fig10}}
\end{figure}

\begin{figure}
\centering
\centerline{\includegraphics[scale=0.6]{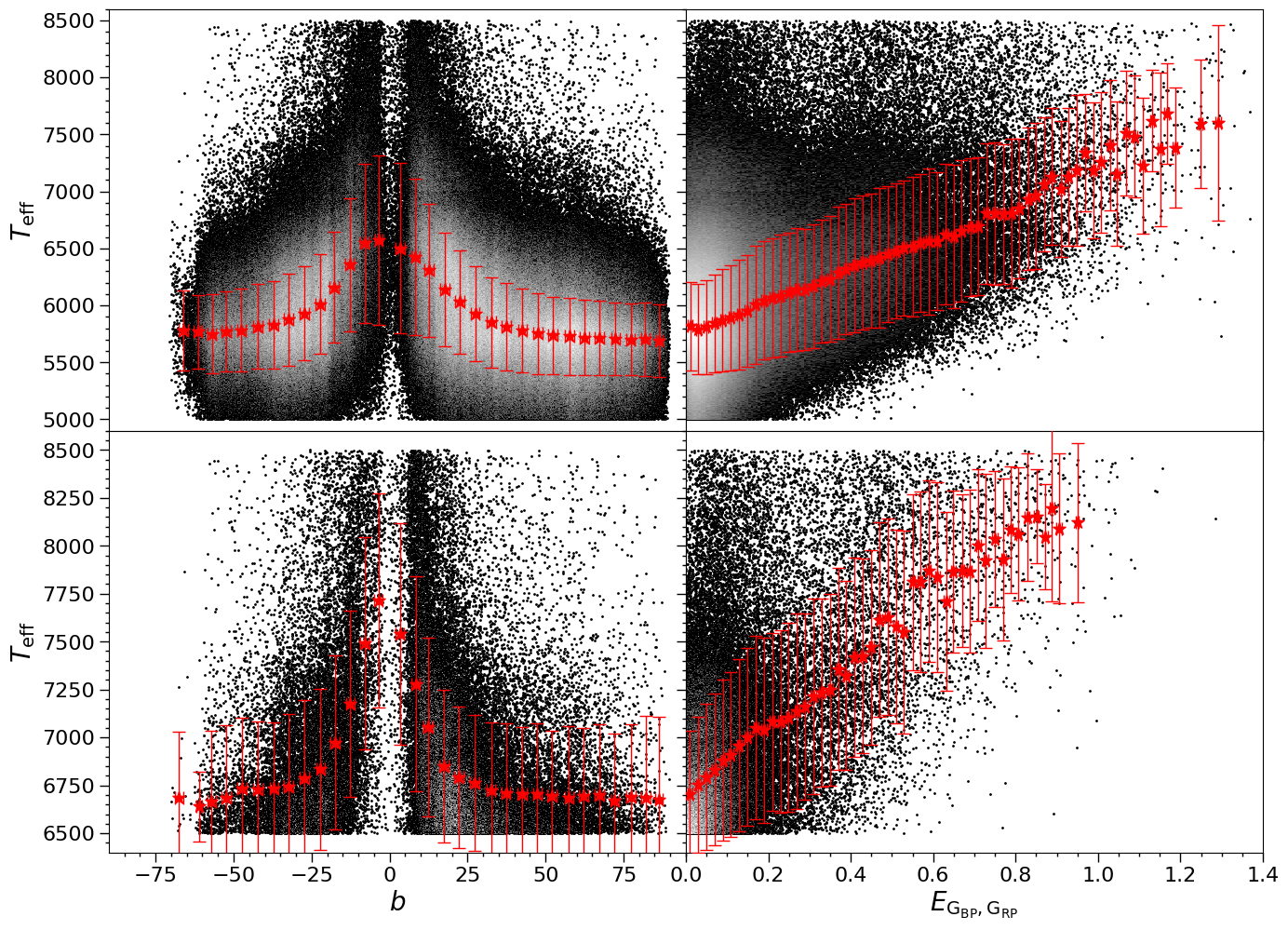}}
\caption{The change of $\Teff$ with Galactic latitude (left panel) and $E_{\rm G_{BP},G_{RP}}$ (right panel). The upper panel is the result of NUV catalog, The lower panel is the result of FUV catalog. The gray-scale encodes the source density. The red asterisks denote the median value for every $5^{\circ}$ in the Galactic latitude (left panel) and every 0.02 mag in $E_{\rm G_{BP},G_{RP}}$ (right panel). The red error bar shows the standard deviation.
\label{fig11}}
\end{figure}

Comparing these expectations with what is found in Figure \ref{fig11}, it can be argued that the tendency of the change of  $E_{\rm NUV,G_{BP}}/E_{\rm G_{BP},G_{RP}}$  cannot be accounted for by stellar $\Teff$ and the color excess. The measured ratio $E_{\rm NUV,G_{BP}}/E_{\rm G_{BP},G_{RP}}$ shows no increasing tendency with $E_{\rm G_{BP},G_{RP}}$, moreover, a small rise appears at small $E_{\rm G_{BP},G_{RP}}$, which disagrees with the increasing trend of  from $\sim$ 3.1 at low extinction to $\sim$ 3.2 at high extinction expected from by Figure \ref{fig10} and Figure \ref{fig11}. For $E_{\rm FUV,G_{BP}}/E_{\rm G_{BP},G_{RP}}$ which is basically constant, there is no evident contradiction with the expectation brought by the difference of $\Teff$ and $E_{\rm G_{BP},G_{RP}}$. Nevertheless, the small decrease at the very low $E_{\rm G_{BP},G_{RP}}$ is inconsistent with the expectation from Figure \ref{fig10} and Figure \ref{fig11}, i.e. from $\sim$ 3.0 at low extinction to $\sim$ 3.2 at high extinction.  We may conclude that the change of $E_{\rm NUV,G_{BP}}/E_{\rm G_{BP},G_{RP}}$ and $E_{\rm FUV,G_{BP}}/E_{\rm G_{BP},G_{RP}}$ needs some variation of the dust properties to account for, such as the dust size distribution, but we will not go further on this point.

\subsection{The near ultraviolet extinction map}

\begin{figure}
\centering
\centerline{\includegraphics[scale=0.5]{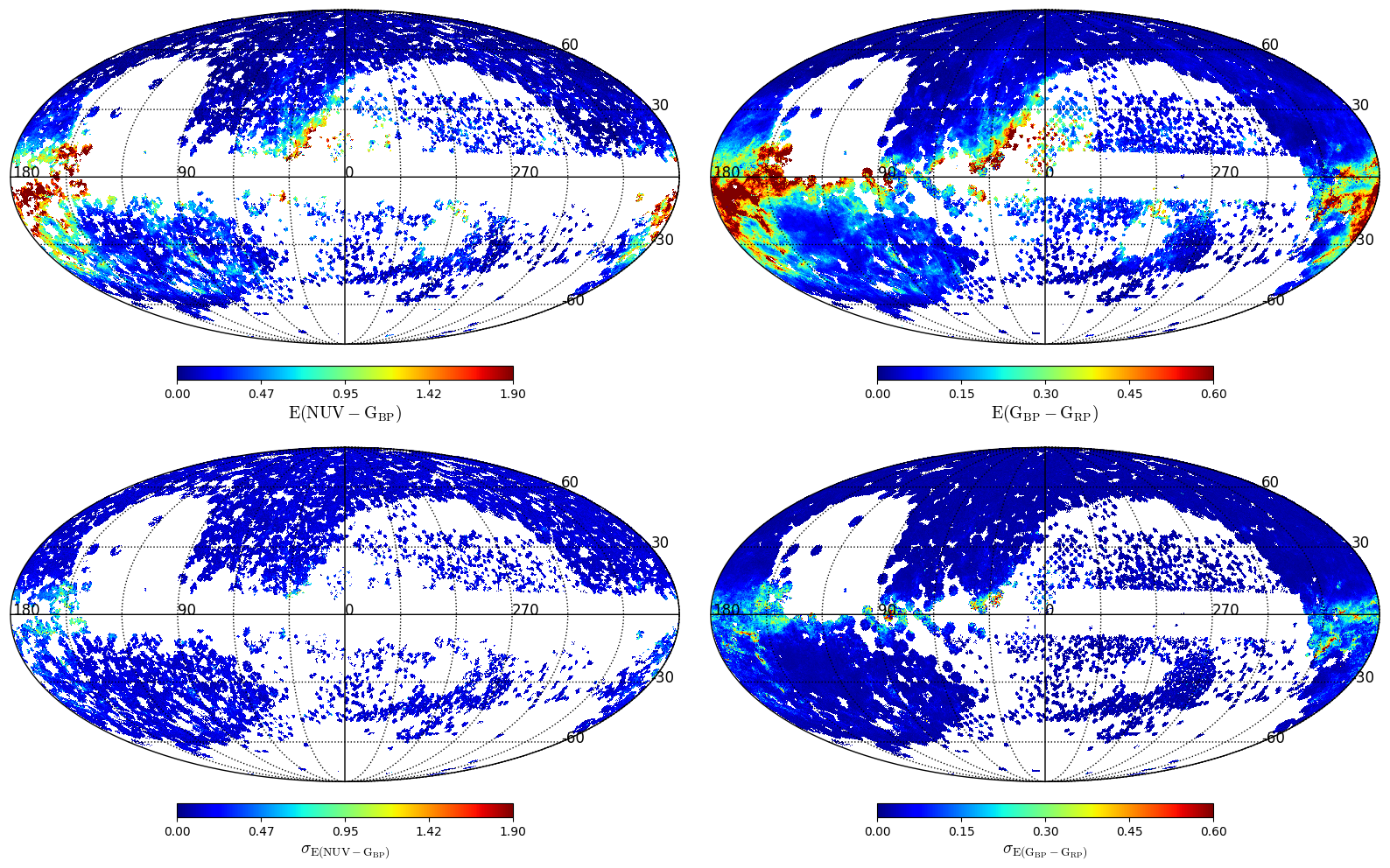}}
\caption{Gridding map by the HEALPix method of $ E_{{\rm NUV,G_{BP}}}$ (upper left), $E_{{\rm G_{BP},G_{RP}}}$ (upper right) and their errors (lower panels).
\label{fig12}}
\end{figure}

\begin{figure}
\centering
\centerline{\includegraphics[scale=0.6]{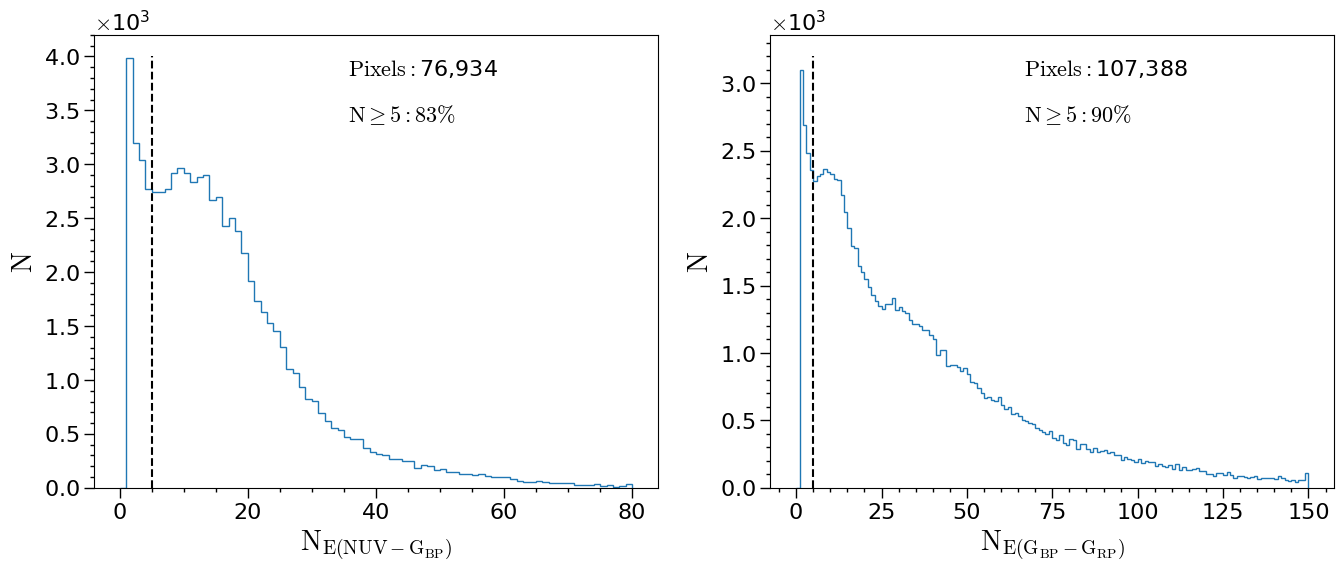}}
\caption{The histogram of the number of sources in each HEALPix pixel for  $ E_{\rm NUV,G_{BP}}$ (left) and $E_{\rm G_{BP},G_{RP}}$ (right).
\label{fig13}}
\end{figure}

With the color excess $E_{{\rm NUV,G_{BP}}}$  for  more than one million stars, the NUV extinction map can be constructed. But for the $FUV$ band, only about 50,000 measurements are available, which is not enough to delineate the FUV extinction sky, so the $FUV$ band will not be studied further. In order to make the NUV extinction map, the HEALPix algorithm (Hierarchical Equal Area isoLatitude Pixelization) for pixelizing a sphere \citep{2005ApJ...622..759G} is used with the python code "healpy" \footnote{\url{https://healpy.readthedocs.io/}} to grid our data. We select $ N_{\rm side}$ = 128 and divide the whole sky into 196,608 pixels ($ N_{\rm pix}$), which means about 0.2 $\rm deg^2$ per pixel. The median value by an iterative 3$\sigma$ clipping is taken as the representative value of each bin.  The map of median values and standard deviations are shown in the left column of Figure~\ref{fig12}. This map has 76,934 pixels covering a third of the sky, 83\% of which has more than 5 sources (see Figure~\ref{fig13}). In comparison, the map of $E_{{\rm G_{BP},G_{RP}}}$ has 107,388 pixels covering half sky, 90\% of which has more than 5 sources (see the right column of Figure~\ref{fig12} and \ref{fig13}). It should be noticed that there are 3,987 pixels in $E_{\rm NUV,G_{BP}}$ and 3,098 pixels in $E_{\rm G_{BP},G_{RP}}$ with only one source which correspondingly yields a zero standard deviation and is not taken into account in the lower panels of Figure \ref{fig12}.

The large scale structures of the \emph{NUV} extinction map agree very well with the optical one, as shown in Figure \ref{fig12}. The most prominent features include the Orion and the Tau-Per-Aur complexes at the anti-center direction, the Ophiuchus cloud at the Galactic center direction, the Aquila South and the Chamaeleon clouds along $b \sim -20^{\circ}$. The general increase towards the Galactic plane is apparent, with some small-scale structures at high latitudes.

\section{Summary}

Using the stellar parameters from the updated spectroscopic database from the LAMOST survey in the northern hemisphere and the GALAH survey in the south, the accurate interstellar extinctions in the\emph{ GALEX} ultraviolet bands are obtained for a significantly increased number of stars. This large sample forms the solid basis to construct the extinction map and to study the extinction law in the ultraviolet. The main results of this work are following: 

\begin{enumerate}

\item In combining the stellar parameters from the LAMOST and GALAH spectroscopic survey with the color indexes from the  \emph{GALEX} and \emph{Gaia} photometric survey,  the color excesses $E_{{\rm NUV,G_{BP}}}$ and  $E_{{\rm FUV,G_{BP}}}$ are obtained for 1,244,504 stars and 56,123 stars by using the blue-edge method to calculate the intrinsic color indexes. For reference,  the color excess $E_{{\rm G_{BP},G_{RP}}}$ is obtained  for  4,169,769 stars.

\item  The color excess ratios, $ E_{{\rm NUV,G_{BP}}} / E_{{\rm G_{BP},G_{RP}}}$ and $ E_{{\rm FUV,G_{BP}}} / E_{{\rm G_{BP},G_{RP}}}$, are derived to be $3.25 \pm 0.04$ and $2.95 \pm 0.16$ respectively for the entire sample, which highly agrees with the derived $ E_{{\rm FUV,NUV}} / E_{{\rm G_{BP},G_{RP}}}= -0.38 \pm 0.24$  as well as with our previous result  \citep{2018ApJ...861..153S}. The derived $E_{{\rm FUV,NUV}} / E_{{\rm G_{BP},G_{RP}}}$ at $ |b|>20^{\circ}$ is also negative.

\item The UV extinction map is constructed by using the HEALPix method ($N_{\rm side}=128$), which covers about a third of the sky with an angular resolution of $\sim 0.4\,\, \rm deg$. Correspondingly, the optical extinction map covers about half of the sky.

\end{enumerate}

\acknowledgments{We thank Prof. Jian Gao, and Mr. Yi Ren, Bin Yu and  Ye Wang for helpful discussions.  We thank the referee for his/her very helpful suggestion.  This work is supported by National Key R\&D Program of China No. 2019YFA0405503, the CSST Milky Way Survey on Dust and Extinction Project and NSFC 11533002.  This work made use of the data taken by \emph{GALEX}, LAMOST, GALAH, and \emph{Gaia}.
}

\software{Astropy \citep{2013A&A...558A..33A},
Dustmap \citep{2018JOSS....3..695M},
Healpy \citep{Zonca2019},
Topcat \citep{2005ASPC..347...29T}}

\facilities{\emph{GALEX}, LAMOST, GALAH, \emph{Gaia}}

\clearpage
\bibliographystyle{aasjournal}
\bibliography{ccm1}


\begin{deluxetable}{rrrrrrrrrrr}
\caption{\label{table1} The sources whose $E_{\rm B,V}^{Green}$ is very small but $E_{\rm G_{BP},G_{RP}}$ is greater than 2 mag}
\tablehead{\colhead{${\rm Gaia\, EDR3\,source\_id}$} & \colhead{$l$} & \colhead{$b$} & \colhead{$E_{\rm G_{BP},G_{RP}}$} & \colhead{$E_{\rm B,V}^{Green}$} & \colhead{$T_{\rm eff}$} & \colhead{Z} & \colhead{$log$g} & \colhead{${\rm G_{BP}}$} & \colhead{${\rm G_{RP}}$} & \colhead{Distance}\\
		\colhead{$$}&\colhead{$(^{\circ})$}& \colhead{$(^{\circ})$}& \colhead{$(mag)$} & \colhead{$(mag)$}& \colhead{$(K)$} & \colhead{$(dex)$} & \colhead{$(dex)$} & \colhead{$(mag)$} & \colhead{$(mag)$} & \colhead{$(pc)$}
}
\startdata
$677542433954708352$ & 199.5102 & 27.3609 & 2.047 & 0.000 & 5333.100 & -0.401 & 4.012 & 18.262 & 15.294 & 103.347 \\
$4470223578258953984$ & 31.5775 & 11.1872 & 2.149 & 0.000 & 5445.190 & -0.563 & 4.061 & 18.009 & 14.994 & 161.916 \\
$4471811754082004864$ & 32.9590 & 12.2938 & 2.022 & 0.000 & 5526.600 & -0.505 & 4.163 & 18.263 & 15.399 & 196.890 \\
$615947655126410496$ & 221.0884 & 45.6911 & 2.239 & 0.000 & 5510.750 & -0.317 & 4.406 & 15.956 & 12.855 & 0.000 \\
$638858934828614784$ & 205.8427 & 42.0043 & 2.195 & 0.000 & 5631.910 & -0.621 & 4.168 & 17.716 & 14.721 & 78.050 \\
$612073663345730560$ & 210.2949 & 36.6923 & 2.069 & 0.000 & 5867.070 & -0.946 & 4.088 & 14.159 & 11.359 & 0.000 \\
$3380587660834921344$ & 192.4363 & 12.5757 & 2.569 & 0.020 & 5820.660 & 0.096 & 4.352 & 11.473 & 8.110 & 1739.761 \\
$3152555992983369472$ & 210.3120 & 6.2716 & 2.011 & 0.000 & 5929.570 & -0.108 & 4.192 & 18.279 & 15.516 & 101.379 \\
$3351670367587793792$ & 201.9155 & 5.3588 & 2.239 & 0.000 & 6064.320 & 0.156 & 4.219 & 18.221 & 15.253 & 122.230 \\
$357101005674463232$ & 133.6188 & -13.3826 & 2.557 & 0.000 & 6116.290 & -0.245 & 4.235 & 18.597 & 15.341 & 194.236 \\
$298863379442524544$ & 137.3431 & -33.4778 & 2.162 & 0.000 & 6237.350 & -0.502 & 4.207 & 17.330 & 14.508 & 157.496 \\
$942904303983636864$ & 177.3206 & 11.2839 & 2.028 & 0.000 & 6214.500 & -0.313 & 4.355 & 18.162 & 15.462 & 205.876 \\
$896972484905743872$ & 183.0170 & 22.7020 & 2.136 & 0.000 & 6330.760 & -0.286 & 4.153 & 16.969 & 14.187 & 79.970 \\
$243224139711234944$ & 154.0198 & -9.7273 & 2.000 & 0.000 & 6426.810 & -0.423 & 4.279 & 18.611 & 15.991 & 195.610 \\
$159957467005754752$ & 171.9277 & -8.6219 & 2.242 & 0.000 & 6455.300 & 0.027 & 4.134 & 17.170 & 14.304 & 89.379 \\
$965547307143587584$ & 169.3304 & 15.8453 & 2.274 & 0.000 & 6552.730 & -0.478 & 4.218 & 18.564 & 15.702 & 187.668 \\
$163222814678059648$ & 168.1840 & -16.5454 & 2.523 & 0.000 & 6532.650 & 0.215 & 4.288 & 17.022 & 13.892 & 164.081 \\
$3329089250611733376$ & 199.9983 & -4.0750 & 2.187 & 0.000 & 6977.620 & -0.142 & 4.035 & 18.240 & 15.562 & 189.627 \\
$3369765575904237056$ & 194.4654 & 2.8603 & 2.263 & 0.000 & 6971.350 & -0.025 & 4.031 & 18.286 & 15.528 & 141.350 \\
$1871304232836895360$ & 79.6792 & -6.3520 & 2.160 & 0.008 & 7420.670 & -0.047 & 4.418 & 18.463 & 15.922 & 351.710 \\
$3378872761998931200$ & 192.7499 & 7.5942 & 2.194 & 0.000 & 7707.700 & -0.374 & 4.118 & 18.453 & 15.932 & 345.223
\enddata
\end{deluxetable}

\clearpage
\end{CJK*}
\end{document}